%% file: main.tex
\documentclass[aps,prl,reprint,superscriptaddress,nobibnotes]{revtex4-2}

\usepackage{amsmath,amssymb,bm}
\usepackage{graphicx}
\usepackage{xcolor}
\usepackage{microtype}
\usepackage{hyperref}
\usepackage{orcidlink}
\usepackage{multirow}

\newcommand{\dd}{\mathrm{d}}
\newcommand{\Rop}{\mathcal{R}}

\newcommand{\ketPhi}{\lvert\Phi\rangle}
\newcommand{\braPhi}{\langle\Phi\rvert}
\newcommand{\Ne}{$^{20}$Ne}
\newcommand{\Mg}{$^{34}$Mg}
\newcommand{\Mgt}{$^{32}$Mg}
\newcommand{\Texas}{N$^3$LO$_{\mathrm{Texas}}$}

\begin{document}

\title{Interpretation follows computation for deformed atomic nuclei}

\author{G.~Hagen\orcidlink{0000-0001-6019-1687}}
\affiliation{Physics Division, Oak Ridge National Laboratory, Oak Ridge,
Tennessee 37831, USA}
\affiliation{Department of Physics and Astronomy, University of Tennessee,
Knoxville, Tennessee 37996, USA}
\affiliation{Department of Physics and Astronomy, Chalmers University of Technology,
  SE-412 96 G\"oteborg, Sweden}

\author{T.~Papenbrock\orcidlink{0000-0001-8733-2849}}
\affiliation{Department of Physics and Astronomy, University of Tennessee,
Knoxville, Tennessee 37996, USA}
\affiliation{Physics Division, Oak Ridge National Laboratory, Oak Ridge,
Tennessee 37831, USA}

\date{\today}

\begin{abstract} 
  The interpretation of nuclear levels in experiments and ab initio
  computations is challenging because it requires the production and
  analysis of many expectation values and transition strengths.
  Here we introduce an approach that yields directly interpretable
  states. Our ab initio coupled-cluster computations start from a reference
  state with axial symmetry and positive parity. Equation-of-motion
  techniques then yield intrinsic states with definite parity
  and angular-momentum projection onto the symmetry axis. Finally, the
  projection onto good angular momenta yields rotational bands based
  on these intrinsic states. In the neutron-rich nucleus $^{34}$Mg,
  where only the ground-state band is known, we find several low-lying 
  positive- and negative-parity bands. The lowest, a
  $K^\pi=3^-$ band, is dominated by a single neutron particle-hole
  excitation rather than by octupole collectivity. These results
  challenge the common wisdom about how deformation and pairing
  work in nuclei.
\end{abstract}

\maketitle

\textit{Introduction.---} The low-lying states of deformed heavy
nuclei can be organized into rotational
bands~\cite{bohr1975,ring1980}.  Each band is thought to consist of
the rotational states associated with a deformed and correlated
intrinsic state that exhibits at most two quantum numbers: parity
($\pi$) and, in the case of axially symmetric intrinsic states, the
projection $K$ of the total angular momentum onto the symmetry axis.
While the Nilsson model~\cite{nilsson1955} provides us with an
interpretation of such states within a single-particle model, the
actual identification of the $K^\pi$ quantum numbers and the
placement of states into a rotational band requires much more than
the energy levels. In experiment, one needs to measure associated
gamma-ray transitions and use their intensities to place levels into
bands~\cite{davidson1981,aprahamian2006}. The situation is similarly complicated for some ab
initio~\cite{hergert2020,ekstrom2023} methods. The no-core shell model, for
instance, identifies rotational-band states with angular momentum $J$
by their characteristic $J(J+1)-K^2$ pattern (for $K\ne 1/2$), their
electric quadrupole moments, and their $B(E2)$ transition
strengths~\cite{caprio2013,caprio2015} or, based on Elliott's SU(3)
symmetry~\cite{elliott1958}, by selecting basis states with the
empirical SU(3) patterns for their
computation~\cite{dytrych2013,launey2016}. Other ab initio
computations~\cite{yao2020,frosini2022b,sun2025,papenbrock2026} followed the path
charted by nuclear density functional
methods~\cite{egido1991,rodriguez2002,paar2007,niksic2011,shen2019,sheikh2021}
and computed rotational bands that resulted from deformed (and
symmetry-breaking) ground states after symmetry projection. This
approach retains the direct interpretability of the computed
states. However, the computed states are limited to those that
result from constrained mean-field computations and the corresponding
collective coordinates (such as quadrupole, octupole, or triaxial
deformations).

This paper has two goals. First, we considerably enlarge the set of computed states
with a direct interpretation by also allowing for non-collective
excitations of the reference state. We will use coupled-cluster equation-of-motion
methods~\cite{sekino1984,geertsen1989,stanton1993,comeau1993,kowalski2004,kowalski2004b,piecuch2005,gour2006,wloch2006, bartlett2007, piecuch2009, jansen2011,hagen2014} to compute excitations of the correlated and axially symmetric ground state. We focus on excitations that are dominated by
one-particle--one-hole (1p-1h) excitations (but will also include
2p-2h excitations). Such excitations include collective
excitations (that can be reached by constrained mean-field states),
because a change in the mean field is dominated by a coherent
superposition of 1p-1h excitations. However, our approach also reaches
non-collective states, which are dominated by a small number of 1p-1h
excitations and can therefore be readily interpreted as well. So far, such
approaches have been limited to odd and odd-odd nuclei by simply
having the odd nucleons blocking certain single-particle states~\cite{banerjee1974,mang1975,bally2014,borrajo2018,robledo2019,bally2021,sun2025b}. Here, however, we focus on even-even nuclei, where such blocking techniques
fail.

\begin{figure*}[!thb]
 \centering
 \includegraphics[width=\textwidth]
 {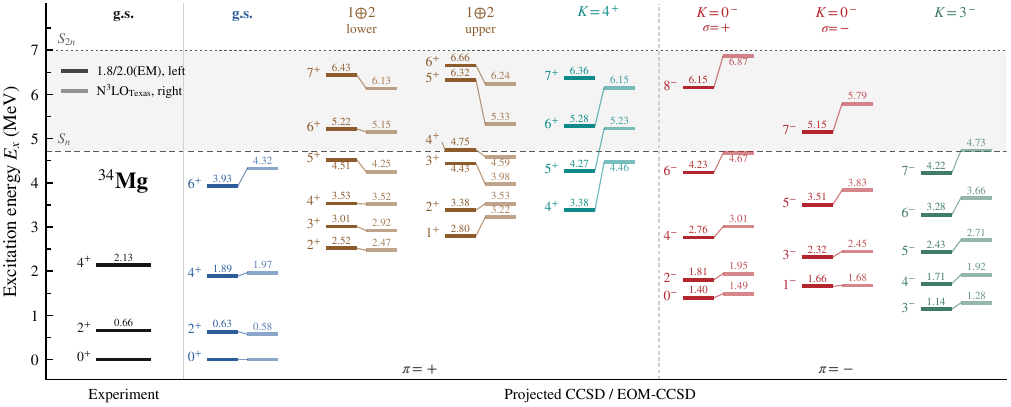}
 \caption{Predicted spectrum of \Mg{} compared with the experimentally
   known ground-state band~\cite{michimasa2014,ensdf}. Computed rotational
   bands (left levels: 1.8/2.0~(EM) interaction~\cite{hebeler2011}; right levels: \Texas{} interaction~\cite{hu2026}) are grouped by parity and labeled by the $K^\pi$ quantum
   numbers of the underlying intrinsic state(s). The two $K^\pi=0^-$
   bands are the signature partners of one intrinsic state.  The two positive-parity columns labeled $1\oplus2$ are
   the eigenstate sequences of the intrinsic $K^\pi=1^+$
   and $2^+$ states that mix under rotation (End Matter).
   Shading marks the region between the one- and two-neutron
   separation energies, $S_n=4.7$ and
   $S_{2n}=7.0$~MeV~\cite{wang2021}, respectively.}
 \label{fig:mg34}
\end{figure*}

The second goal of this paper is to report the existence of several negative-parity bands at lowest energy in the exotic nucleus $^{34}$Mg. This nucleus is in the
so-called ``island of inversion''~\cite{warburton1990}, i.e., neutron-rich
isotopes of neon, sodium, and magnesium with neutron number $N\ge
20$ that are deformed in their ground
states~\cite{campi1975,wildenthal1980,poves1987}. The deformation is
easily understood in the Nilsson model, which also suggests that 1p-1h
excitations could populate negative-parity states in even-even nuclei
at very low energies. We indeed find that this is the case. This nucleus defies the textbook picture of the Bohr
Hamiltonian, where $K^\pi=0^+$ and $2^+$ are the quantum numbers of
the lowest rotational bands in deformed nuclei (because these result
from collective quadrupole deformations of the nuclear surface). Some
actinide and rare-earth nuclei are already known to break with this picture by exhibiting collective octupole-deformed states
that are the band heads of the lowest-lying rotational
bands~\cite{egido1991,butler1996}. In contrast, and as we will see
below, the low-lying negative-parity bands in $^{34}$Mg are not
collective octupole excitations of the ground state. So, $^{34}$Mg is
deformed, but it is incorrect to think about its lowest-energy
excitations as the surface deformations and oscillations of a liquid drop. We also note
that negative-parity excitations in even-even nuclei require the breakup of a fermion pair in time-reversed orbitals.  Combining these last two points
with the low-lying negative-parity bands suggests to us that $^{34}$Mg
is probably too light to exhibit the typical properties of a deformed and superfluid liquid
drop that is so prevalent for most open-shell nuclei.

These exciting results are shown in Fig.~\ref{fig:mg34}. We computed
intrinsic states with $K^\pi=0^\pm, 1^\pm, \ldots,4^\pm$ and kept
those whose energies are considerably below the neutron separation
thresholds. The shown rotational bands result from projection of
angular momentum using the established
techniques~\cite{duguet2015,qiu2017,qiu2018,hagen2022,sun2025}. 
We used two different interactions to make sure that our results are robust. In what follows we describe how we arrived at these results.

\textit{Hamiltonian and model space.---}
We start from the intrinsic Hamiltonian
\begin{equation}
  \label{Ham}
  H=T-T_{\mathrm{CM}}+V_{\mathrm{2N}}+V_{\mathrm{3N}} \ .
\end{equation}
Here, $T$ and $T_{\mathrm{CM}}$ are the total kinetic energy and that
of the center of mass, respectively. We use two different sets of two-nucleon and three-nucleon potentials $V_{\mathrm{2N}}$ and $V_{\mathrm{3N}}$, namely the interaction 1.8/2.0~(EM)~\cite{hebeler2011} and \Texas{}~\cite{hu2026}. Both interactions are from chiral effective field theory. The 1.8/2.0~(EM) interaction is
accurate for binding energies and spectra across the nuclear
chart~\cite{hagen2017,morris2018,bonaiti2025}; the most recent \Texas{} interaction is expected to be also accurate for very neutron-rich nuclei. 
We perform our calculations in a spherical harmonic-oscillator basis
with $e\equiv 2n+\ell\leq e_{\max}$ and $\hbar\omega=16$~MeV, using
the symmetry-conserving two-body rank reduction of the three-nucleon
force of Ref.~\cite{frosini2021} obtained from a spherical equal-filling Hartree--Fock density. We use $e_{\max}=8$ in what follows, with three-body
matrix elements truncated at $E_{3\max}=16\hbar\omega$. Starting from this rotationally invariant Hamiltonian we construct the deformed Hartree--Fock basis from which the ensuing coupled-cluster computations start. A Hartree--Fock
computation then yields an axially symmetric, prolately deformed
reference state $\ketPhi$, which conserves parity and has good angular
momentum projection $M=0$ along the laboratory $z$ axis. For the
reference state, this axis is also the intrinsic symmetry axis, and we
can therefore identify $M$ with $K$. The axially symmetric reference
state simplifies the coupled-cluster computations and
the angular-momentum projection. Normal ordering with respect to the
reference state $\ketPhi$ yields the normal-ordered Hamiltonian $H_N$.

\textit{Coupled-cluster computations.---} We perform coupled-cluster with
singles and doubles (CCSD)
computations~\cite{kuemmel1978,bartlett2007,hagen2014}. This yields
the similarity-transformed Hamiltonian
\begin{equation}
  \label{Hsim}
\overline{H}=e^{-T} H_N e^T \ .
\end{equation}
Here $T=T_1+T_2$ consists of the 1p-1h and 2p-2h excitation
operators. These preserve $J_z$ and parity. One can think of
$e^T\ketPhi$ as the intrinsic correlated ground state; it has $M=0$
and positive parity. We next compute excited states using
equation-of-motion coupled-cluster with singles and doubles
(EOM-CCSD).  The biorthogonal EOM-CCSD
states are
\begin{equation}
 \lvert\Psi_\nu^R\rangle = R_\nu e^T\ketPhi,\qquad
 \langle\Psi_\nu^L\rvert = \braPhi L_\nu e^{-T} \ .
 \label{eq:eom-states}
\end{equation}
Here $R_\nu$ and $L_\nu$ consist of 1p-1h and 2p-2h excitation and
de-excitation operators, respectively. The index $\nu=(K,\pi,\omega)$ labels an EOM-CCSD state by its axial quantum numbers $K_\nu^\pi$ and by its excitation energy $\omega_\nu$. Biorthogonality means
$\langle\Psi^L_\mu\vert\Psi^R_\nu\rangle=\delta_{\mu\nu}$. We solve
Eq.~\eqref{eq:eom-states} by finding the right and left eigenvectors
of the similarity-transformed Hamiltonian~(\ref{Hsim}) in the space of
allowed (i.e., having the targeted $K$ and $\pi$ quantum numbers)
particle-hole excitations. Table~\ref{tab:cr} shows the excitation energies $\omega_\nu$ of the computed states with quantum numbers $K_\nu^\pi$ resulting from EOM-CCSD, for both interactions. Also shown is the contribution $n_{1p1h}$ of the 1p-1h amplitudes to the norm of each state. We note here that EOM-CCSD also yields spurious states. Those are center-of-mass excitations (which have $K^\pi=0^-$ and $K^\pi=1^-$) or rotations of the deformed ground state (which have $K^\pi=1^+$). Those are easily eliminated (see End Matter) and not reported in Table~\ref{tab:cr}.

\begin{table}[t]
 \caption{\label{tab:cr}
 Results for the intrinsic EOM-CCSD states $\nu$ of \Mg{}, labeled by $K_\nu^\pi$, whose rotational bands are shown in Fig.~\ref{fig:mg34}.    
 $n_{1p1h}$ is the percentage of the biorthogonal norm carried
 by 1p-1h excitations; $\omega_\nu$ is the EOM-CCSD excitation energy; $\delta_\nu$ is the energy gain from triples excitations; $\delta E_\nu$ is the energy gain of the band head from symmetry projection; $E_{\rm bh}=\omega_\nu-\delta_\nu+\delta E_{\rm g.s.}
 -\delta E_\nu$ is the
 excitation energy of the band head 
 relative to the projected $0^+$ ground state (g.s.). The subscripts $\sigma\pm$ give the signature of the two $K^\pi=0^-$ states. For $K^\pi=1^+$ and $2^+$, Fig.~\ref{fig:mg34} shows the mixed projection.
 EM denotes the 1.8/2.0~(EM) interaction and TX the \Texas{} interaction. The $n_{1p1h}$ values are those of EM and agree with TX within one percentage point.
 All energies are in MeV.}
 \begin{ruledtabular}
 \begin{tabular}{lccccccccc}
 & & \multicolumn{2}{c}{$\omega_\nu$} & \multicolumn{2}{c}{$\delta_\nu$} & \multicolumn{2}{c}{$\delta E_\nu$} & \multicolumn{2}{c}{$E_{\rm bh}$}\\
 $K_\nu^\pi$ & $n_{1p1h}$ (\%) & EM & TX & EM & TX & EM & TX & EM & TX\\
 \hline
 $0^+_{\rm g.s.}$ & -- & $0$ & $0$ & -- & -- & $1.69$ & $2.35$ & $0$ & $0$\\
 $1^+$ & $90$ & $3.14$ & $3.79$ & $1.03$ & $1.14$ & $1.01$ & $1.78$ & $2.78$ & $3.22$\\
 $2^+$ & $91$ & $3.10$ & $2.92$ & $0.99$ & $1.12$ & $1.00$ & $1.68$ & $2.80$ & $2.47$\\
 $4^+$ & $89$ & $4.01$ & $4.89$ & $1.21$ & $1.29$ & $1.11$ & $1.49$ & $3.38$ & $4.46$\\
 $0^-_{\sigma+}$ & $90$ & $1.95$ & $2.23$ & $0.86$ & $1.35$ & $1.38$ & $1.74$ & $1.40$ & $1.49$\\
 $0^-_{\sigma-}$ & $90$ & $2.11$ & $2.35$ & $0.91$ & $1.41$ & $1.22$ & $1.62$ & $1.66$ & $1.68$\\
 $3^-$ & $90$ & $1.50$ & $1.83$ & $0.82$ & $1.33$ & $1.22$ & $1.57$ & $1.14$ & $1.28$\\
 \end{tabular}
 \end{ruledtabular}
  \end{table}

The energies of the deformed ground and excited states computed so far are still not accurate enough. While we keep their wave functions at the 2p-2h level, energy corrections from triples (i.e., 3p-3h excitations) can be included  with the
state-specific, noniterative CR-EOMCC(2,3),A correction of Refs.~\cite{kowalski2004,wloch2006,piecuch2005,piecuch2009}; such corrections were applied to nuclear structure problems in Refs.~\cite{gour2005,wloch2005,binder2013}. They rely on perturbative estimates from 3p-3h amplitudes without requiring their computation or storage. 
The energy gain from triples, $\delta_\nu$, is also shown in Table~\ref{tab:cr}. It is about 1~MeV for 1.8/2.0~(EM) and between $1.1$ and $1.4$~MeV for \Texas{}, and it is clearly state specific.

The final step in our computation is angular-momentum projection. This is done for the CCSD ground state and the EOM-CCSD excitations. Neglecting triples excitations is a good approximation because those mainly introduce short-range correlations into the wave function~\cite{bell2022}, while angular-momentum projection is mainly concerned with the inclusion of long-range correlations~\cite{hagen2022,sun2025}.  
The projected left and right EOM-CCSD states with angular-momentum projection $M$ in the laboratory system are 
 \begin{equation}
  \lvert\Psi^{JM}_\nu\rangle=P^J_{MK_\nu}\lvert\Psi^R_\nu\rangle,\quad
  \langle\widetilde{\Psi}^{JM}_\nu\rvert=\langle\Psi^L_\nu\rvert P^J_{K_\nu M},  \label{eq:projected-states}
\end{equation}
with $(P^J_{MK})^\dagger=P^J_{KM}$. 
Our EOM-CCSD states are eigenstates of $J_z$ with eigenvalue $K_\nu$, and the integral over the third Euler angle $\gamma$ in $P^J_{MK}$ contains $\int\dd\gamma e^{i(K-K_\nu)\gamma}$, which vanishes unless $K=K_\nu$. In the norm and Hamiltonian kernels the product $P^J_{KM}P^J_{MK'}=P^J_{KK'}$ does not depend on $M$, and the two azimuthal Euler-angle integrations leave the projection operator
\begin{equation}
 P^J_{KK'}=\frac{2J+1}{2}\int_0^\pi\dd\beta\,\sin\beta\,
 d^J_{KK'}(\beta)\,\Rop(\beta),
 \label{eq:projector}
\end{equation}
with $\Rop(\beta)=e^{-i\beta J_y}$. The norm and Hamiltonian kernels
between the EOM states $\mu$ and $\nu$,
\begin{equation}
 N^J_{\mu\nu}=\langle\Psi_\mu^L\rvert P^J_{K_\mu K_\nu}\lvert\Psi_\nu^R\rangle,\quad
 H^J_{\mu\nu}=\langle\Psi_\mu^L\rvert P^J_{K_\mu K_\nu}H\lvert\Psi_\nu^R\rangle,
 \label{eq:kernels}
\end{equation}
enter the generalized eigenvalue problem
$\sum_\nu H^J_{\mu\nu}f^J_\nu=E^J\sum_\nu N^J_{\mu\nu}f^J_\nu$ for the
states $\lvert\Psi^{JM}\rangle=\sum_\nu f^J_\nu P^J_{MK_\nu}\lvert\Psi^R_\nu\rangle$. This is similar to the generator-coordinate method~\cite{hill1953,griffin1957}.
Most bands shown in Fig.~\ref{fig:mg34} are projections of a single EOM state, i.e., the kernels are diagonal and we have 
$E^J_\nu=H^J_{\nu\nu}/N^J_{\nu\nu}$. 
However, some of the intrinsic EOM-CCSD states are connected by a rotation. Thus, they mix in the angular-momentum projection. In collective models of the nucleus, this mixing is due to the Coriolis coupling. 

Each Hartree--Fock orbital of the axial reference state has a definite
projection $m$ of its angular momentum on the symmetry axis (which is the $z$ axis), and we write
$\Omega=\lvert m\rvert$ for the Nilsson label of the time-reversed pair
with $m=\pm\Omega$.  A particle-hole pair with
$\Omega_p$ for the particle and $\Omega_h$ for the hole therefore gives two EOM states, namely
$K=\Omega_p+\Omega_h$ and $K=\lvert\Omega_p-\Omega_h\rvert$, depending on
which member of the hole pair is emptied.  The two states differ by
$2\min(\Omega_p,\Omega_h)$ units of $K$.  The rotation operator in Eq.~\eqref{eq:projector} is
$\Rop(\beta)=e^{-i\beta J_y}=1-\tfrac{\beta}{2}(J_+-J_-)+\mathcal O(\beta^2)$,
and its first-order term changes $K$ by one unit.  For $\Omega_h=1/2$
the two couplings of a pair differ by one unit of $K$ and are connected
already at this order.  In the particle-rotor model this is the Coriolis
coupling~\cite{kerman1956,bohr1975,papenbrock2020}.  
In this case the single-state
projections are not eigenstates and we diagonalize $H$ in the space of
both $K$-values. For $\Omega_h\geq3/2$ the two couplings differ by
three or more units of $K$ and are connected only at third or higher
order in $\beta$, and we keep their single-state projections. We note that the even-$J$ positive-parity states with $K^\pi=1^+$ and $2^+$ carry a small admixture of the spurious rotated copy of the ground-state band, and we therefore include this state in the mixing so that the
diagonalization returns the copy of the ground-state band as a separate
solution, which we discard (see End Matter).

The key technical step is to evaluate the angle-dependent kernels
$\langle\Psi_\mu^L\rvert\Rop(\beta)\lvert\Psi_\nu^R\rangle$ and
$\langle\Psi_\mu^L\rvert\Rop(\beta)H\lvert\Psi_\nu^R\rangle$.
For the ground 
state this is solved by the disentangled
formulation~\cite{qiu2017,qiu2018,hagen2022,sun2025}, which exploits the Thouless theorem~\cite{thouless1960}. For a rotation with nonzero overlap $\langle\Phi\rvert\Rop(\beta)\lvert\Phi\rangle$ we have
\begin{equation}
\braPhi\Rop(\beta)=\langle\Phi\rvert\Rop(\beta)\lvert\Phi\rangle\braPhi e^{V(\beta)} \ ,
 \label{eq:thouless}
\end{equation}
where $V(\beta)$ is a one-body de-excitation operator~\cite{qiu2017}. 
Acting to the right on the correlated coupled-cluster state introduces the product
$e^{V}e^{T}\ketPhi$, which one writes in the disentangled approach as
\begin{equation}
 e^{V(\beta)}e^{T}\ketPhi
 =e^{W_0(\beta)+W_1(\beta)+W_2(\beta)+\cdots}\ketPhi \ .
 \label{eq:disentangle}
\end{equation}
Here $W_0$ is a number and $W_n$ for $n\geq1$ is an $n$p-$n$h excitation operator, and we truncate after $W_2$. This is an accurate approximation (see Supplemental Material). To solve for $W_n$ we integrate a set of ordinary differential equations~\cite{qiu2017, bally2018,sun2025}. To apply this approach to excited states we note that $T$ and $R_\nu$ commute. Thus, one can write
\begin{equation}
 R_\nu e^T\ketPhi
 =\left.\partial_\eta
 e^{T+\eta R_\nu}\ketPhi\right|_{\eta=0} \ .
 \label{eq:tangent}
\end{equation}
The exponent $T+\eta R_\nu$ is formally of the same structure as $T$, and the disentangled approach can be utilized also here. 
The derivative is evaluated as a centered finite difference of the kernels with a small $\eta$ (End Matter).
The angular-momentum projection splits the deformed intrinsic states into rotational bands. For the band head, the energy gain from angular-momentum projection, $\delta E_\nu$, is also shown in Table~\ref{tab:cr} together with the final result for the excitation energy of the band head, $E_\mathrm{bh}$. All members of a band carry the energy gain from triples $\delta_\nu$ of their
intrinsic state. For a state obtained from the mixed projection, the
energy gain from triples is the average of the $\delta_\nu$ of the
physical intrinsic states that enter, weighted by their mixture in
that state.
We see that the net gain from angular-momentum projection, i.e., the difference $\delta E_\mathrm{g.s.}-\delta E_\nu$, is just a few hundred keV, and that the ground state gains most from the projection. Because of this, the energy correction from projection cancels some of the small energy gains from triples $\delta_\nu$, and the EOM-CCSD results ($\omega_\nu$) are just about 0.5~MeV above the band-head energy $E_\mathrm{bh}$. In agreement with simple estimates from an effective theory for deformed nuclei~\cite{hagen2022,coelloperez2025}, the energy gain from angular-momentum projection is larger for the ground-state band because it has the smaller moment of inertia. With \Texas{} the energy gain from triples and the projection gain of the ground state are both larger than with 1.8/2.0~(EM), and the two effects again largely cancel in the band-head energies. 

\textit{Discussion of Figure~\ref{fig:mg34}.---}
From experiments only the ground-state band is known in $^{34}$Mg~\cite{church2005,michimasa2014,doornenbal2013}, and measurements
in neighboring isotopes documented the competing
configurations~\cite{motobayashi1995,wimmer2010,crawford2016}.
There are a few theoretical results. Ab initio calculations with the 1.8/2.0~(EM) interaction used the valence-space in-medium similarity renormalization group (VS-IMSRG) in multishell valence spaces~\cite{miyagi2020,cao2025} or the projected generator
coordinate method~\cite{zhou2025}. They only computed positive-parity states.
\textcite{ohta2004} employed a nuclear density functional and used parity projection of a symmetry-broken mean-field state. For $^{34}$Mg, they reported results for the odd-spin $K^\pi=0^-$ band~\cite{ohta2005}. They placed the $J^\pi=1^-$ state about $1$~MeV above ours, and their $7^-$ state is at about 9~MeV. They also concluded that this band does not come from a collective octupole excitation.  

Let us focus on our results. We note that the computed states from both interactions are close to each other, and their difference is a measure of the Hamiltonian uncertainty. The band heads of the $K^\pi=3^-$ and $0^-$ bands in Table~\ref{tab:cr} agree within $0.15$~MeV between the two interactions, and those of the positive-parity bands within $1.1$~MeV. We also note that the results are readily interpreted within the Nilsson diagram~\cite{nilsson1955,ring1980}. For the reference
state obtained with the 1.8/2.0~(EM) interaction the $\Omega^\pi=1/2^-$ orbital of the $f_{7/2}$ shell has crossed
below the $\Omega^\pi=3/2^+$ orbital of the $d_{3/2}$ shell. Thus, the 22 neutrons fill the $sd$ shell and the $\Omega^\pi=1/2^-$ orbital, and a deformed gap separates the occupied $\Omega^\pi=3/2^+$ from the empty $\Omega^\pi=3/2^-$ orbital of the
$f_{7/2}$ shell. The lowest empty positive-parity orbital lies
much higher and low-lying excitations are therefore dominated by neutrons. With \Texas{} we obtain a different reference state. Here the $\Omega^\pi=3/2^-$ orbital is the last occupied orbital and the $\Omega^\pi=3/2^+$ orbital is empty. For both interactions, the $K^\pi=3^-$ and $0^-$ states have the same intrinsic configuration, i.e., one neutron in each of the $3/2^-$ and $3/2^+$ orbitals (Supplemental Material~\cite{suppmat}), and the $K^\pi=3^-$ and $0^-$ bands are very similar. 
The $K^\pi=3^-$ state is lower in energy than the $K^\pi=0^-$ states, in agreement with the phenomenological Gallagher rule~\cite{gallagher1962} (Supplemental Material~\cite{suppmat}). The $K^\pi=0^-$ bands are based on two intrinsic (time-reversed) states, with one neutron in the $m^\pi= \pm3/2^+$ and 
$m^\pi=\mp3/2^-$ orbitals. Angular-momentum projection couples these excitations and yields two signature bands with even and odd values of angular momentum $J$. The odd-$J$ members lie about $0.1$~MeV above the $J(J+1)$ curve of the even-$J$ members, a displacement analogous to the Newby shift in $K=0$ bands of odd-odd nuclei~\cite{newby1962} (Supplemental Material~\cite{suppmat}).
We note that the single configuration with one neutron in each of the $\Omega^\pi=3/2^+$ and $\Omega^\pi=3/2^-$ orbitals carries more than 90\% of the weight of all 1p-1h excitations that make the intrinsic $K^\pi=0^-$ and $3^-$ states. Thus, the intrinsic states are very simple when viewed in the deformed Hartree--Fock basis. Clearly, these are the relevant single-particle degrees of freedom.

The positive-parity EOM states are excitations within a major shell. For the 1.8/2.0~(EM) interaction, the $K^\pi=4^+$ state is the $K=\Omega_p+\Omega_h$ coupling of the proton $3/2^+\to5/2^+$ excitation within $d_{5/2}$. The $K^\pi=1^+$ state carries the $K=\lvert\Omega_p-\Omega_h\rvert$ coupling of the same excitation with 68\% of its weight and the neutron $1/2^-\to3/2^-$ excitation within $f_{7/2}$ with 28\%. 
The $K^\pi=2^+$ state is the $K=\Omega_p+\Omega_h$ coupling of the neutron excitation. Its hole has $\Omega_h=1/2$, so the $K^\pi=1^+$ and $K^\pi=2^+$ couplings are connected by the Coriolis coupling, and the projected $K^\pi=1^+$ and $2^+$ states mix almost equally into the two sequences of Fig.~\ref{fig:mg34}.
With \Texas{} the $K^\pi=4^+$ state is instead the neutron $3/2^-\to5/2^-$ excitation within $f_{7/2}$, and the
$K^\pi=2^+$ and $1^+$ states are the antiparallel and parallel couplings, respectively, of the neutron $1/2^+\to3/2^+$
excitation ($s_{1/2}\to d_{3/2}$).  They barely mix, and the two sequences of Fig.~\ref{fig:mg34}
keep at least $97\%$ of a single $K$.

\textit{Summary and outlook.---}
We predicted that negative-parity rotational bands are the low-energy excitations of the deformed ground state of \Mg{}. This is a bit unexpected in view of traditional nuclear structure. Our ab initio computations are conceptually simple and based on an intrinsic deformed state, its particle-hole excitations, and their symmetry projection. In our approach, interpretation follows computation (similar to architectural paradigms about form and function). While the computed spectra are similar for two calibrated interactions, the corresponding states have different configurations. It will be interesting to confront these results with data, e.g. $B(E2)$ transition strengths. As an outlook, one could further increase the number of interpretable states by starting from different (e.g., shape-coexistent) reference states  or by starting from a triaxially deformed reference state. While such computations and developments are left for the future, it is clear that the path proposed in this paper significantly increases the number of states that can be computed ab initio while still being directly interpretable.

\begin{acknowledgments}

G.H. acknowledges support and
hospitality from Chalmers University of Technology through the
Chalmers Jubilee Professorship. 
This work was supported by the U.S. Department of Energy, Office of
Science, Office of Nuclear Physics, under Award No.~DE-FG02-96ER40963,
and by SciDAC-5 (NUCLEI collaboration). 
Computer time was provided by the Innovative and Novel
Computational Impact on Theory and Experiment (INCITE) program.  This
research used resources of the Oak Ridge Leadership Computing Facility
at Oak Ridge National Laboratory, supported by the Office of Science
of the Department of Energy under Contract No.~DE-AC05-00OR22725. 
The authors acknowledge the use of Claude (Anthropic) for assistance with code development and debugging, proofreading, checking quoted numbers, and figure layout, and of ChatGPT (OpenAI) for improving the clarity of the manuscript. All scientific results and conclusions were independently verified by the authors.

\end{acknowledgments}

\input{main.bbl}

\clearpage

\input{endmatter}

\clearpage

\input{supplement}

\end{document}

%% file: main.bbl
%

%% file: endmatter.tex
\section*{End Matter}




\subsection*{Angular momentum projection of excited states}

The angle-dependent kernels entering Eq.~\eqref{eq:kernels} are
\begin{align}
\mathcal N_{\mu\nu}(\beta)&=\langle\Psi_\mu^L\rvert\Rop(\beta)\lvert\Psi_\nu^R\rangle \ , \nonumber\\
\mathcal H_{\mu\nu}(\beta)&=\langle\Psi_\mu^L\rvert\Rop(\beta)H\lvert\Psi_\nu^R\rangle \ .  
\end{align}
They are computed using the expressions
\begin{align}
 \mathcal N_{\mu\nu}(\beta)&=n(\beta)\,\partial_\eta
 \Bigl[e^{W_0}\braPhi\widetilde Z_\mu\,e^{W}\ketPhi\Bigr]_{\eta=0},\nonumber\\
 \mathcal H_{\mu\nu}(\beta)&=n(\beta)\,\partial_\eta
 \Bigl[e^{W_0}\braPhi\widetilde Z_\mu\,H_V\,e^{W}\ketPhi\Bigr]_{\eta=0} \ .
 \label{eq:kernels-eta}
\end{align}
Here $n(\beta)=\braPhi\Rop(\beta)\ketPhi$,
$\widetilde Z_\mu=e^{V}\Rop^{-1}Z_\mu\Rop e^{-V}$, and
$H_V=e^{V}He^{-V}$.  All quantities in the brackets depend on $\beta$,
and $W_0$ and $W$ also on $\eta$.  We suppress these arguments for
notational simplicity.  The left EOM state
$\braPhi Z_\mu=\braPhi L_\mu e^{-T}$ and its one-body transform
$\braPhi\widetilde Z_\mu$ contain at most 2h-2p de-excitations.
The dependence on the state $\nu$ enters through the disentangled
amplitudes, defined by
\begin{equation}
 e^{V(\beta)}e^{T+\eta R_\nu}\ketPhi
 =e^{W_0(\beta;\eta)}e^{W(\beta;\eta)}\ketPhi .
 \label{eq:disentangle-eta}
\end{equation}
Here $W_0$ is a scalar and $W=W_1+W_2+\cdots$ contains particle-hole
excitations.  At $\eta=0$ this reduces to the ground-state
disentangling, Eq.~\eqref{eq:disentangle}.  At each fixed angle $\beta$,
we replace $V(\beta)$ with $\lambda V(\beta)$ and integrate the resulting
ODEs from $\lambda=0$ to $1$, starting from $W_0=\eta r_0$,
$W_1=T_1+\eta R_{\nu,1}$, and $W_2=T_2+\eta R_{\nu,2}$, where the scalar
$r_0$ of $R_\nu$ vanishes except for $K^\pi=0^+$.
The integration runs over $\lambda$ for fixed angle $\beta$.

We turn to the evaluation of the derivatives in Eq.~(\ref{eq:kernels-eta}). Let us denote the expressions in the square brackets as $F(\eta)$.  For the derivative of $F(\eta)$ we evaluate it at $\eta=\pm\varepsilon$, with
$\varepsilon=10^{-4}$. (This yields a small error of order $O(\varepsilon^2)$.)  We checked that the kernels remain stable
for $10^{-2}\geq\varepsilon\geq3\times10^{-5}$.  The transformed left EOM
state $\braPhi\widetilde Z_\mu$ is independent of $\eta$ and is computed
once per angle. We note that $F(\eta)$ is odd in $\eta$ for negative-parity EOM states, because the negative-parity $Z_\mu$ connects only terms with an odd number of $R_\nu$. Thus, $F(0)=0$ and $F(-\eta)=-F(\eta)$, and the centered difference reduces to $F(\varepsilon)/\varepsilon$ and comes at the cost of a single evaluation. 


For spin-0 ground states, the kernels~(\ref{eq:kernels-eta}) are symmetric  under $\beta\to \pi-\beta$~\cite{bohr1975}, and this allows us to limit the rotation angle to the interval $0\leq\beta\leq\pi/2$.  Kernels for excited states in general lack  this
symmetry, and we use $0\leq\beta\leq\pi$ for all excited states.

Because the disentangling is truncated at $W_2$ in Eq.~(\ref{eq:disentangle-eta}), the projected states are not exact eigenstates of $\hat{J}^2$. The deviation of $\langle J^2\rangle$
from $J(J+1)$ measures that error directly.   Table~\ref{tab:j2} reports this
diagnostic for the two $K^\pi=0^-$ signature partners of \Mg{} at
$e_{\max}=6$.  Overall, the projections are quite accurate. We also show the contribution $N_\nu^J$ of each angular momentum $J$ to each of the intrinsic states with a definite signature. 

\begin{table}[t]
 \caption{\label{tab:j2}Projected $\langle J^2\rangle$ and projected
 norms $N^J_\nu$ for the two $K^\pi=0^-$ signature partners of \Mg{} at
 $e_{\max}=6$ with a $20$-point mesh.}
 \begin{ruledtabular}
 \begin{tabular}{lccc}
 $J^\pi$ & $N^J_\nu$ & $\langle J^2\rangle$ &
 $\langle J^2\rangle-J(J+1)$\\
 \hline
 \multicolumn{4}{l}{Positive signature ($\sigma=+1$)}\\
 $0^-$  & $0.095$ & $0.109$   & $\phantom{-}0.109$\\
 $2^-$  & $0.357$ & $6.077$   & $\phantom{-}0.077$\\
 $4^-$  & $0.328$ & $20.064$  & $\phantom{-}0.064$\\
 $6^-$  & $0.163$ & $42.147$  & $\phantom{-}0.147$\\
 $8^-$  & $0.048$ & $72.011$  & $\phantom{-}0.011$\\
 \hline
 \multicolumn{4}{l}{Negative signature ($\sigma=-1$)}\\
 $1^-$  & $0.257$ & $2.094$   & $\phantom{-}0.094$\\
 $3^-$  & $0.374$ & $12.063$  & $\phantom{-}0.063$\\
 $5^-$  & $0.248$ & $30.096$  & $\phantom{-}0.096$\\
 $7^-$  & $0.095$ & $56.162$  & $\phantom{-}0.162$\\
 \end{tabular}
 \end{ruledtabular}
\end{table}


\subsection*{Validation in \Ne{}}

\begin{figure*}[htb]
 \centering
 \includegraphics[width=0.96\textwidth]{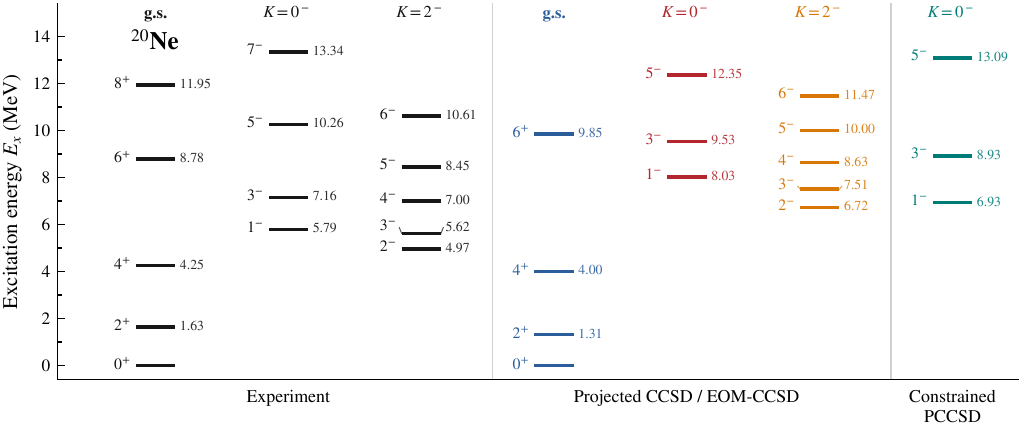}
 \caption{\label{fig:ne20}Spectrum of \Ne{}.  The experimental ground
   band and the evaluated $K^\pi=0^-$ and $2^-$
   bands~\cite{tilley1998} (left) are compared with the results from
   this work (middle) projected CCSD ground band and those from
   Ref.~\cite{hagen2026a} (right).  Spin-parity ($J^\pi$) and energies
   (in MeV) are printed next to the levels.}
\end{figure*}

\Ne{} is a textbook case of a prolate nucleus with low-lying
$K^\pi=0^-$ and $2^-$ bands, headed by $J^\pi=1^-$ and $2^-$ states at
$5.79$ and $4.97$~MeV, respectively~\cite{tilley1998}.  Its intrinsic
and cluster structure has been probed
experimentally~\cite{nauruzbayev2017,mehl2025}, and described by
theory~\cite{rodriguezguzman2003,frosini2022b,heller2023}.
\textcite{kanadaenyo2020}  computed both negative-parity bands, and their mixing, with
antisymmetrized molecular dynamics. They found that the 
$K^\pi=0^-$ band is associated with octupole and $\alpha$-cluster
degrees of freedom~\cite{zhou2016,marevic2018,sargsyan2025}.
\textcite{frosini2022b,hagen2026a} computed this band from
octupole-constrained reference states, with each level taken at the
lowest projected energy along the constraint. In the present work, a
parity-conserving prolate Hartree--Fock minimum is the reference
state, and the two bands are obtained from the projections of the
$K^\pi=0^-$ and $K^\pi=2^-$ EOM states.  Figure~\ref{fig:ne20} shows
these results (together with the positive-parity ground-state band) and compares
them to data.  The $K^\pi=0^-$ band is also compared to a recent
computation based on an octupole-deformed reference
state~\cite{hagen2026a}.  
The projected CCSD ground-state band is close
to experiment, but resembles more a rigid-rotor spectrum than the
data. This is probably due to our approach based on projection after
variation.  

The computed negative-parity bands are a bit too high in
energy. This is consistent with the view that the corresponding
intrinsic state can be viewed as a $\alpha+{}^{16}$O cluster
state~\cite{kanadaenyo2020,zhou2016,marevic2018,sargsyan2025}.
Inclusion of such 4p-4h excitations is beyond our computational means
(but would presumably lower the energy of the intrinsic state). 
Overall, we view the results presented in Fig.~\ref{fig:ne20} as a validation of our approach used in \Mg{}.

\subsection*{Elimination of spurious states}

The Hamiltonian~(\ref{Ham}) references neither the center-of-mass
coordinate $\mathbf{R}=A^{-1}\sum_{j=1}^A\mathbf{r}_j$ nor its canonical momentum, i.e. the total
momentum.  The coupled-cluster wave function therefore factorizes
approximately into an intrinsic part and a center-of-mass Gaussian whose
frequency $\tilde\omega$ differs from the oscillator frequency
$\omega$~\cite{hagen2009a}.  For an axially symmetric reference state the
Gaussian has two frequencies, $\tilde\omega_z$ and $\tilde\omega_\perp$.
The center-of-mass wave function can be excited by the one-body operator
$\mathbf{R}$ or by its conjugate momentum  (which is the generator of translations); this yields spurious excitations. Were the factorization between the center-of-mass and intrinsic wavefunction exact, the spurious states would have zero energy. Alas, the factorization is only approximate and the EOM spectrum can contain spurious center-of-mass excitations at
small but nonzero energy.  To
identify them we compute the biorthogonal transition strength of the
center-of-mass dipole operator $Q_{1M}\propto\sum_ir_iY_{1M}(\hat{\bm r}_i)$,
with $M=0,\pm 1$, between the CCSD ground state and each EOM state.  The dipole operator is a
rank-one spherical tensor and can connect the $0^+$ ground state only to negative-parity
states with $|M|\leq 1$, i.e. $K^\pi=0^-$ and $1^-$. Thus, bands with $K\geq 2$ cannot carry this
strength.   In \Mg{}, the lowest $K^\pi=0^-$ and $1^-$ states, at
$0.085$ and $0.574$~MeV, respectively, have transition matrix elements with= $|Q_{10}|^2=15.97$~fm$^2$ and
$|Q_{11}|^2=13.18$~fm$^2$, respectively, and are excluded.  In contrast, the physical $K^\pi=1^-$
state has $|Q_{11}|^2=1.2\times10^{-5}$~fm$^2$, six orders of magnitude
smaller. We note that the $n^{\rm th}$ power $\mathbf{R}^n$ can also create spurious center-of-mass excitations. However, this is a $n$-body operator and it
generates states with significant $n$-particle--$n$-hole content. As we limit ourselves to states with dominant 1p-1h excitations, such spurious states are of no concern to us.

The intrinsic reference states breaks rotational symmetry. Spurious rotational excitations then stem from Nambu-Goldstone modes and are generated by $\hat{J}_{\pm 1}$ operators. Thus EOM states with  $K^\pi=1^+$ could be such spurious states~\cite{petrik2018}. The lowest
$K=1^+$ EOM state of \Mg{}, at $1.69$~MeV, carries
$|\langle L_\nu\vert J_+\vert0\rangle|^2=16.3$ (and is thereby identified as spurious), compared with $0.015$
for the physical $K^\pi=1^+$ state.  The projected norm of the spurious $K=1^+$ EOM state lies
predominantly in even $J$, as expected for a rotated copy of the
even-$J$ ground band~\cite{bohr1975}, and it has no $1^+$ bandhead. For the 1.8/2.0~(EM) interaction the physical $K^\pi=1^+$ and $2^+$ states are built on the same 1p-1h excitations as
the spurious state.  For even $J$ their projections are therefore not
orthogonal to the projection of the spurious state. Including the spurious state, the generalized eigenvalue problem yields the copy of the ground-state band as one of its solutions. The other solutions are biorthogonal to it and thereby free of the admixture.  The odd-$J$ states are not affected because the ground-state band has no odd-$J$ members.

%% file: supplement.tex
\section{Supplemental Material}

This Supplemental Material presents the particle-hole content of the
intrinsic EOM-CCSD states of \Mg{} for the 1.8/2.0~(EM) and \Texas{}
interactions, and projected EOM-CCSD results for \Mgt{}, the $N=20$
neighbor of \Mg{} in which negative-parity states have been observed
(but not yet ordered into bands).  All calculations use the model
space and methods of the main text, with $e_{\max}=8$,
$E_{3\max}=16\hbar\omega$, and $\hbar\omega=16$~MeV. 
The \Mgt{}
results are based on the 1.8/2.0~(EM) interaction.

\subsection*{Particle--hole content of the \Mg{} bands}
\begin{figure*}[t]
 \centering
 \includegraphics[width=0.92\textwidth]{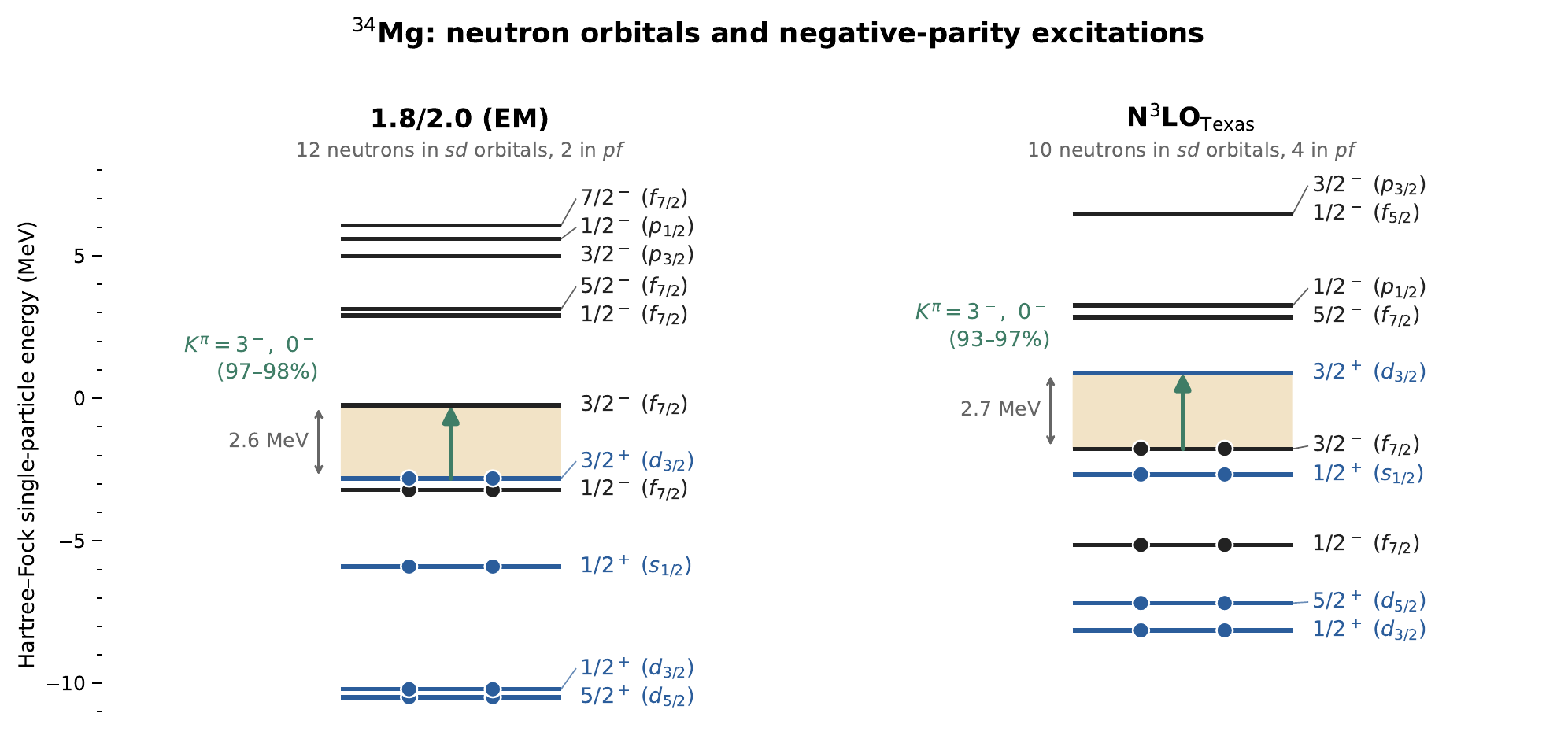}
 \caption{\label{fig:mg34refs}Neutron Hartree--Fock orbitals of the \Mg{} reference states obtained with the
 1.8/2.0~(EM) interaction (left) and with \Texas{} (right). Orbitals are labeled by $\Omega^\pi$ with the
 dominant spherical parentage in parentheses. Positive-parity orbitals are blue, negative-parity orbitals are
 black, and dots mark occupied orbitals. The shaded band lies between the last occupied and the
 first empty orbital. The arrow shows the excitation that dominates the $K^\pi=3^-$ and $0^-$ states, with the
 fraction of $n_{1p1h}$ it carries.}
\end{figure*}
\begin{figure*}[t]
 \centering
 \includegraphics[width=0.92\textwidth]{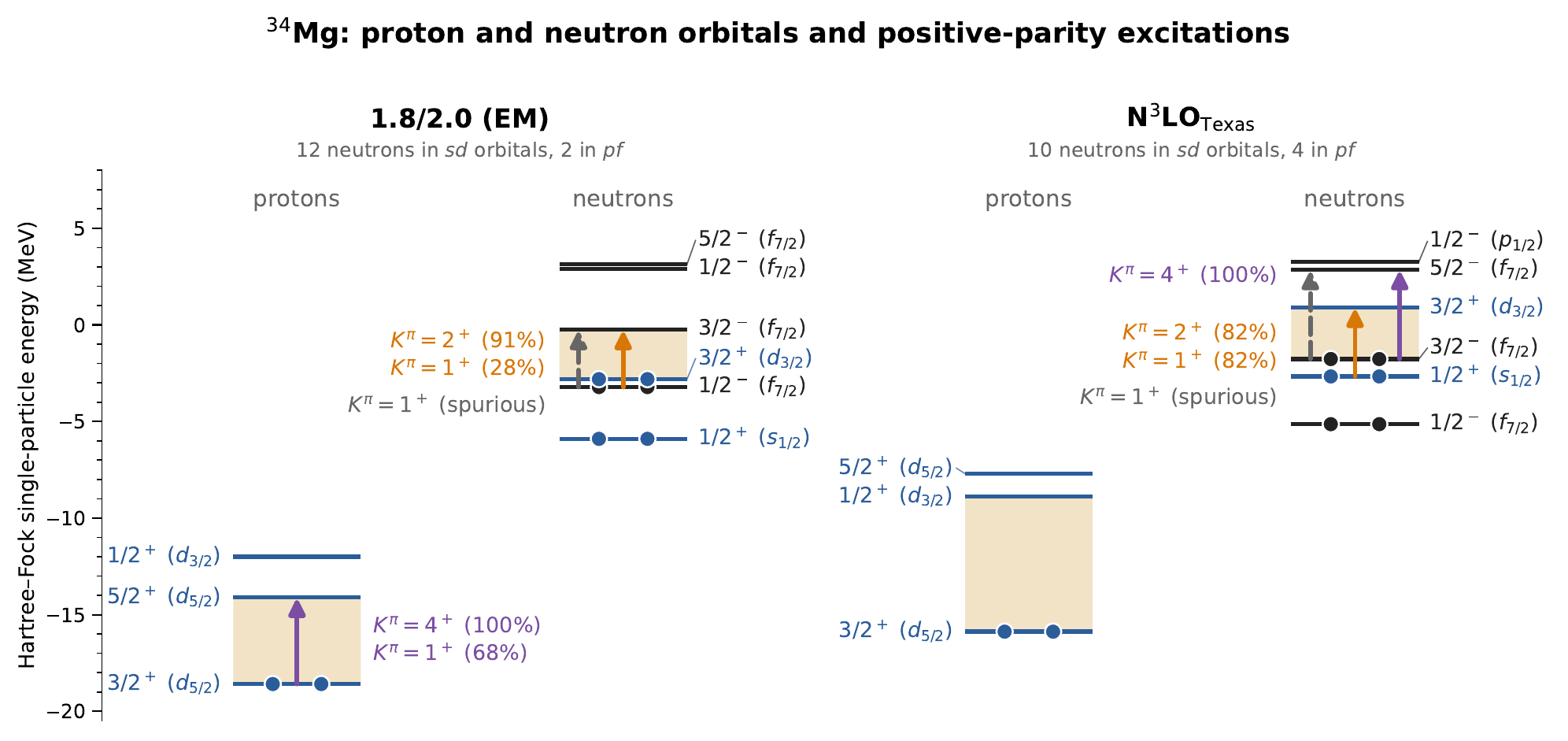}
 \caption{\label{fig:mg34pos}Proton and neutron Hartree--Fock orbitals of \Mg{} and the particle-hole
 excitations that dominate the positive-parity states, for the 1.8/2.0~(EM) interaction (left) and for
 \Texas{} (right). Colors, labels, and shading are as in Fig.~\ref{fig:mg34refs}. Each arrow is labeled with the
 intrinsic states $K^\pi$ it contributes to and with the fraction of $n_{1p1h}$ it carries. The dashed
 arrow is the largest component of the spurious $K^\pi=1^+$ state, the rotation of the ground state.}
\end{figure*}
\begin{table*}[t]
 \caption{\label{tab:ph}Particle--hole content of the intrinsic EOM-CCSD
 states of \Mg{} for the 1.8/2.0~(EM) and \Texas{} (TX) interactions.
 Left: dominant excitations between Hartree--Fock orbitals, labeled by $\Omega^\pi$ with the orbital energy in
 MeV in parentheses, and the fraction of $n_{1p1h}$ 
 they carry, summed over the two time-reversed components for $K=0$.
 Right: the same one-body excitation operator expanded in the spherical
 oscillator basis, as fractions of its squared norm summed over radial and
 magnetic quantum numbers and over both directions of each pair of shells (for positive parity states both protons and neutrons contribute).}
 \begin{ruledtabular}
 \begin{tabular}{llccc@{\hspace{1.2em}}cccccc}
 & & \multicolumn{3}{c}{Hartree--Fock orbitals} & &
 \multicolumn{5}{c}{Spherical $(l,j)$ content}\\
 $K^\pi$ & & hole $\Omega_h^\pi$ ($\varepsilon_h$) & particle $\Omega_p^\pi$ ($\varepsilon_p$) & fraction &
 & $\nu\,d_{3/2}\,f_{7/2}$ & $\nu\,d_{3/2}\,p_{3/2}$ & $\nu\,d_{5/2}\,f_{7/2}$ & &\\
 \hline
 $3^-$ & EM & $3/2^+$ ($-2.81$) & $3/2^-$ ($-0.25$) & $0.98$ & & $0.70$ & $0.16$ & $0.05$ & & \\
 & TX & $3/2^-$ ($-1.77$) & $3/2^+$ ($\phantom{-}0.90$) & $0.97$ & & $0.67$ & $0.15$ & $0.06$ & & \\
 $0^-_{\sigma+}$ & EM & $3/2^+$ ($-2.81$) & $3/2^-$ ($-0.25$) & $0.98$ & & $0.69$ & $0.19$ & $0.04$ & & \\
 & TX & $3/2^-$ ($-1.77$) & $3/2^+$ ($\phantom{-}0.90$) & $0.97$ & & $0.67$ & $0.15$ & $0.07$ & & \\
 $0^-_{\sigma-}$ & EM & $3/2^+$ ($-2.81$) & $3/2^-$ ($-0.25$) & $0.97$ & & $0.74$ & $0.15$ & $0.05$ & & \\
 & TX & $3/2^-$ ($-1.77$) & $3/2^+$ ($\phantom{-}0.90$) & $0.93$ & & $0.65$ & $0.14$ & $0.08$ & & \\
 \hline
 & & & & & & $\pi\,d_{5/2}\,d_{5/2}$ & $\nu\,f_{7/2}\,f_{7/2}$ & $\nu\,f_{7/2}\,p_{3/2}$ & $\nu\,s_{1/2}\,d_{3/2}$ & $\nu\,d_{3/2}\,d_{3/2}$\\
 \hline
 $1^+$ & EM & $\pi\,3/2^+$ ($-18.59$) & $\pi\,5/2^+$ ($-14.11$) & $0.68$ & & $0.63$ & $0.13$ & $0.11$ & $0$ & $0$\\
 &  & $\nu\,1/2^-$ ($-3.22$) & $\nu\,3/2^-$ ($-0.25$) & $0.28$ & &  &  &  &  & \\
 & TX & $\nu\,1/2^+$ ($-2.67$) & $\nu\,3/2^+$ ($\phantom{-}0.90$) & $0.82$ & & $0.01$ & $0.08$ & $0.02$ & $0.49$ & $0.20$\\
 &  & $\nu\,3/2^-$ ($-1.77$) & $\nu\,5/2^-$ ($\phantom{-}2.85$) & $0.10$ & &  &  &  &  & \\
 $2^+$ & EM & $\nu\,1/2^-$ ($-3.22$) & $\nu\,3/2^-$ ($-0.25$) & $0.91$ & & $0.01$ & $0.43$ & $0.34$ & $0$ & $0$\\
 & TX & $\nu\,1/2^+$ ($-2.67$) & $\nu\,3/2^+$ ($\phantom{-}0.90$) & $0.82$ & & $0.02$ & $0.01$ & $0.01$ & $0.51$ & $0.20$\\
 &  & $\pi\,3/2^+$ ($-15.88$) & $\pi\,1/2^+$ ($-8.89$) & $0.11$ & &  &  &  &  & \\
 $4^+$ & EM & $\pi\,3/2^+$ ($-18.59$) & $\pi\,5/2^+$ ($-14.11$) & $1.00$ & & $0.94$ & $0$ & $0$ & $0$ & $0$\\
 & TX & $\nu\,3/2^-$ ($-1.77$) & $\nu\,5/2^-$ ($\phantom{-}2.85$) & $1.00$ & & $0$ & $0.71$ & $0.16$ & $0$ & $0$\\
 \end{tabular}
 \end{ruledtabular}
\end{table*}

Figures~\ref{fig:mg34refs} and \ref{fig:mg34pos} show the Hartree--Fock orbitals of the two
reference states of \Mg{} and the particle-hole excitations that dominate the intrinsic
negative-parity and positive-parity states, respectively. 
We note that both interactions have the highest occupied and lowest unoccupied orbitals exchanged. 

Table~\ref{tab:ph} shows the particle-hole content of the intrinsic
EOM-CCSD states of \Mg{}, in terms of the Hartree--Fock orbitals and in
the spherical oscillator basis. For the discussion of excited states we note that the coupling a particle ($\Omega_p$) and a hole ($\Omega_h$) angular-momentum projection gives $K=\Omega_p+\Omega_h$ and $K=|\Omega_p-\Omega_h|$.

We start with the $K^\pi=3^-$ state and the two
$K^\pi=0^-$ signature partners. 
Though the orbitals directly above and below the Fermi surface are exchanged in the two interactions,  we have identical dominant configurations for the $K^\pi=3^-$ state and the two
$K^\pi=0^-$ signature partners are identical; they consist of a neutron in the $\Omega^\pi=3/2^+$ and $3/2^-$ orbitals.   
In the spherical basis the excitation is mainly $d_{3/2}\to f_{7/2}$
with a $p_{3/2}$ admixture.  In the limit of no interaction the $K^\pi=3^-$ and $K^\pi=0^-$ states
are degenerate.  For two nucleons of the same kind in deformed
orbitals, the Gallagher rule states that the coupling with antiparallel
intrinsic spin projections lies lowest~\cite{gallagher1962}. To interpret the states using this rule, we computed the expectation value of the spin projection $s_z$ in the
two Hartree--Fock orbitals (Table~\ref{tab:sz}).  For both interactions the spin
projections of the particle and the hole are antiparallel in the
$K^\pi=3^-$ state and parallel in the $K^\pi=0^-$ states.  As the
Gallagher rule predicts, the $K^\pi=3^-$ state lies below the
$K^\pi=0^-$ state, by $0.45$~MeV with EM and by $0.40$~MeV with TX.
The splitting is typical of the two-quasiparticle doublets of rare-earth
nuclei~\cite{sood1989}.

\begin{table}[hbt]
 \caption{\label{tab:sz}Spin projections $\langle s_z\rangle$ of the
 hole and the particle in the $K^\pi=3^-$ and $0^-$ states of \Mg{},
 and the EOM-CCSD excitation energies $\omega_\nu$ (MeV).  The hole
 carries $-m$ and $-\langle s_z\rangle$ of the emptied Hartree--Fock
 orbital.  The particle has $m=+3/2$, and the hole has $m=+3/2$ for
 $K^\pi=3^-$ and $m=-3/2$ for $K^\pi=0^-$.  The two energies for
 $K^\pi=0^-$ are those of the $\sigma=+$ and $\sigma=-$ states.}
 \begin{ruledtabular}
 \begin{tabular}{llcccc}
 & $K^\pi$ & $\langle s_z\rangle_h$ & $\langle s_z\rangle_p$ & spins & $\omega_\nu$\\
 \hline
 EM & $3^-$ & $-0.45$ & $+0.39$ & antiparallel & $1.50$\\
    & $0^-$ & $+0.45$ & $+0.39$ & parallel & $1.95$, $2.11$\\
 TX & $3^-$ & $+0.41$ & $-0.47$ & antiparallel & $1.83$\\
    & $0^-$ & $-0.41$ & $-0.47$ & parallel & $2.23$, $2.35$\\
 \end{tabular}
 \end{ruledtabular}
\end{table}

The two $K^\pi=0^-$ configurations are time-reversed images of each other and have the same mean-field energy.  The similarity transformed Hamiltonian, $\overline H$, connects
them, and the EOM states are their symmetric and antisymmetric
combinations, split by $0.155$~MeV ($\sigma=\pm$ in Table~I of the Letter).
A rotation by $\pi$ about an axis perpendicular to the symmetry axis
exchanges the two configurations.  The symmetric ($\sigma=+$) state is
unchanged by this rotation and the antisymmetric ($\sigma=-$) state
changes sign.  As for the ground-state band of an even-even nucleus, the
$\sigma=+$ state therefore contains only even $J$ and the $\sigma=-$
state only odd $J$, and together they form one $K^\pi=0^-$ band with all $J$.
After symmetry projection, and with the triples corrections included,
the even- and odd-$J$ members have the same moment of inertia, but the
odd-$J$ members are displaced upward by $0.09$~MeV relative to the
even-$J$ members.
An analogous displacement of odd-$J$ relative to even-$J$ members occurs in $K=0$ bands of odd-odd nuclei and is known as the Newby shift~\cite{newby1962}.
There the
residual interaction between the odd proton and the odd neutron connects
a $K=0$ configuration with its time-reversed image.  Here the two
time-reversed configurations are particle-hole excitations of a single
neutron, and $\overline H$ connects them in the same way.  In the rotational model it is a parameter fitted to
data, whereas here its size and sign follow from the chiral interaction.

The positive-parity EOM states are excitations within a major shell (Fig.~\ref{fig:mg34pos}). For the 1.8/2.0~(EM) interaction the $K^\pi=4^+$ state is the $K=\Omega_p+\Omega_h$ coupling of the proton $3/2^+\to5/2^+$ excitation within $d_{5/2}$. The $K^\pi=1^+$ state carries the $K=\Omega_p-\Omega_h$ coupling of the same excitation with 68\% of its weight and the neutron $1/2^-\to3/2^-$ excitation within $f_{7/2}$ with 28\%. 
The $K^\pi=2^+$ state is the $K=\Omega_p+\Omega_h$ coupling of the neutron excitation. Its hole has $\Omega_h=1/2$, so the $K^\pi=1^+$ and $K^\pi=2^+$ couplings are connected by the Coriolis coupling, and the projected $K^\pi=1^+$ and $2^+$ states mix almost equally into the two sequences of Fig.~1 of the main text.
With \Texas{} the $K^\pi=4^+$ state is instead the neutron $3/2^-\to5/2^-$ excitation within $f_{7/2}$, and the
$K^\pi=2^+$ and $1^+$ states are the antiparallel and parallel couplings of the neutron $1/2^+\to3/2^+$
excitation ($s_{1/2}\to d_{3/2}$), $0.87$~MeV apart.  They barely mix, and the two sequences of Fig.~1 of the main text
keep at least $97\%$ of a single $K$.

\subsection{Results for \Mgt{}}

Negative-parity states of \Mgt{} have been computed in the shell
model with the SDPF-M~\cite{tripathi2008,kitamura2021,kitamura2022},
SDPF-U-MIX~\cite{caurier2014,kitamura2022}, and
EEdf1~\cite{kitamura2022} interactions, with a valence-space Hamiltonian
derived from the 1.8/2.0~(EM) interaction by the in-medium
similarity renormalization group~\cite{miyagi2020,kitamura2022}, and
with antisymmetrized molecular dynamics~\cite{kimura2002}.
Experimentally, the negative-parity assignments are tentative.
One-neutron knockout from $^{33}$Mg populated a state at $2.859$~MeV
with $J^\pi=(2,3)^-$ and a state at $3.038$~MeV with $J^\pi=(2)^-$, and
a state at $2.551$~MeV is either $1^-$ or
$2^+$~\cite{kitamura2021,kitamura2022}.

\begin{figure*}[t]
 \centering
 \includegraphics[width=0.98\textwidth]{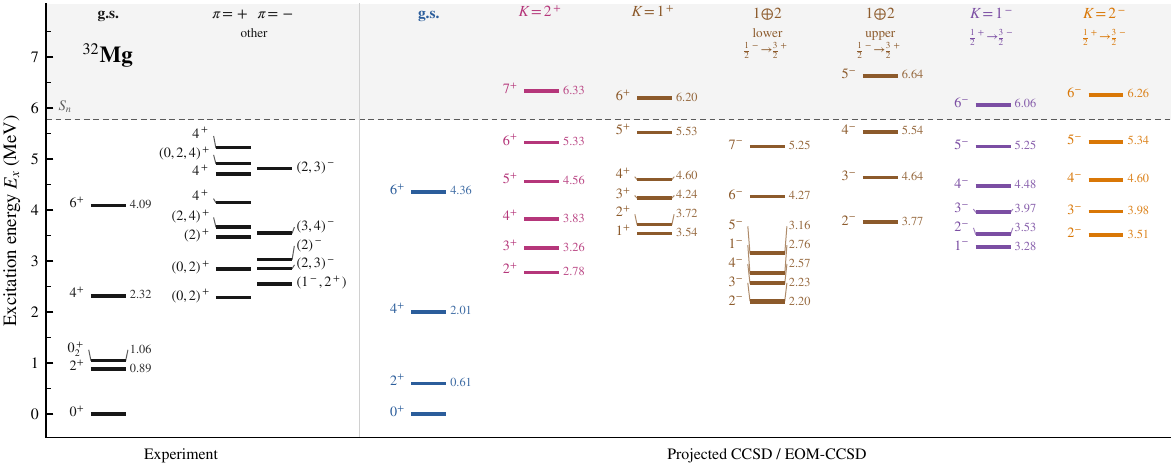}
 \caption{\label{fig:mg32}Spectrum of \Mgt{}.  Experimental results
   (left) are compared to our computations (right).  Energies in MeV
   and spin-parity $J^\pi$ are printed next to the levels.  The
   computed excited states include the triples corrections.  The dashed
   horizontal line marks the neutron-separation energy
   $S_n=5.78$~MeV, and the region above it is shaded.}
\end{figure*}

Figure ~\ref{fig:mg32} shows our projected EOM-CCSD results for \Mgt{} with the 1.8/2.0~(EM) interaction. 
As for $^{34}$Mg we use an axial-symmetric and parity-conserving deformed Hartree--Fock
state as the reference. To make sure we start from the prolately deformed reference and 
not the co-existing near spherical reference state~\cite{sun2025} 
we occupy the last two neutrons at the Fermi surface in the $\Omega^\pi=1/2^-$
orbital of $f_{7/2}$ shell leaving the $\Omega^\pi=3/2^+$ orbital of
$d_{3/2}$ shell is empty. The projected ground-state band is a
bit compressed when compared to data~\cite{ensdf,crawford2016} and a
bit closer to that of a rigid rotor, as was already seen in
Ref.~\cite{sun2025}. That work also computed a second projected $0^+$ 
state based on the near-spherical reference state (i.e. neutrons fully occupy the $d_{3/2}$ shell) a few tens of
keV above the deformed one, consistent with the experimentally observed $0^+_2$ state \cite{wimmer2010}. When viewed from our deformed reference
state the near-spherical configuration cannot be reached by 1p-1h
excitations.

Table~\ref{tab:mg32cr} lists the four lowest negative- and positive parity intrinsic EOM-CCSD states of
\Mgt{} with quantum numbers $K^\pi = 1^\pm, 2^\pm$.  Spurious states of both parities are identified and
eliminated as for \Mg{} (see End Matter).
The negative-parity states fall into two families, family A and B.
Family A moves the intruder neutron back from the $\Omega^\pi=1/2^-$
orbital into the empty $3/2^+$ orbital. As for $^{34}$Mg the intrinsic EOM-CCSD states of \Mgt{} we compute are simple and dominated by 
1p-1h excitations.

\begin{table}[b]
 \caption{\label{tab:mg32cr} Same caption as for Table~I except for \Mgt{}. 
The subscripts A and B denote the
 $1/2^-\to3/2^+$ and $1/2^+\to3/2^-$ families, respectively.}
 \begin{ruledtabular}
 \begin{tabular}{lccccc}
 $K^\pi$ & $n_{1p1h}$ (\%) & $\omega_\nu$ & $\delta_\nu$ & $\delta E_\nu$ & $E_{\rm bh}$\\
 \hline
 $0^+_{\rm g.s.}$ & -- & $0$ & -- & $2.06$ & $0$\\
 $1^+$ & $91$ & $3.79$ & $0.95$ & $1.36$ & $3.54$\\
 $2^+$ & $91$ & $2.98$ & $0.86$ & $1.40$ & $2.78$\\
 $1^-_{\rm A}$ & $90$ & $2.65$ & $1.16$ & $0.80$ & $2.77$\\
 $1^-_{\rm B}$ & $91$ & $3.55$ & $0.81$ & $1.54$ & $3.28$\\
 $2^-_{\rm A}$ & $90$ & $2.41$ & $1.15$ & $0.73$ & $2.59$\\
 $2^-_{\rm B}$ & $90$ & $3.89$ & $0.85$ & $1.59$ & $3.51$\\
 \end{tabular}
 \end{ruledtabular}
\end{table}

As in \Mg{}, we include the Coriolis coupling only where it enters at
first order and connects nearly degenerate intrinsic states (see the
main text).  
This is the case for the $K^\pi=1^-_{\rm A}$ and
$2^-_{\rm A}$ states.  They are the two couplings
$K=\lvert\Omega_p-\Omega_h\rvert=1$ and $K=\Omega_p+\Omega_h=2$ of the
neutron excitation from the $\Omega_h^\pi=1/2^-$ orbital to the
$\Omega_p^\pi=3/2^+$ orbital, and they are split by $0.24$~MeV.  We diagonalize
$H$ in the space of their projections, and Fig.~\ref{fig:mg32} shows the
two resulting sequences with triples contributions added as for $^{34}$Mg. The two states of
family~B and the positive-parity states with $K^\pi=1^+$ and $2^+$ also
differ by one unit of $K$.  The Coriolis coupling of their $1/2^+$ hole
orbital is about ten times weaker than that of the $1/2^-$ orbital and
their intrinsic splittings are larger ($0.34$ and $0.81$~MeV), and we
keep their single-state projections.

As for \Mg{}, we projected the $K^\pi=1^+$ state together with the
spurious rotation of the ground state.  
In contrast to \Mg{}, however, this changes the energies of the states shown in
Fig.~\ref{fig:mg32} only by about $0.01$~MeV, and the $K^\pi=1^+$
state keeps at least $98\%$ of its weight. The $K^\pi=2^+$ state is the $K=2$ coupling of
the neutron $1/2^+\,(s_{1/2})\to3/2^+\,(d_{3/2})$ excitation within the
$sd$ shell, which carries a fraction $0.81$ of its 1p-1h norm, and
the $1^+$ state is the $K=1$ coupling of the same pair (with $0.58$ 1p-1h norm) and with
$0.19$ fraction of the 1p-1h norm coming from the $1/2^-\to3/2^-$ excitation.